\documentclass[12pt]{iopart}

\usepackage{graphicx}
\usepackage{epsfig}

\begin{document}


\textwidth 21cm \advance \textwidth by -5cm \textheight 29.7cm
\advance \textheight by -6cm \oddsidemargin 0cm
\setlength{\oddsidemargin}{0.5cm} \evensidemargin 0cm
\setlength{\evensidemargin}{-0.5cm} \topmargin -0.75cm

%
%
\def\e{\begin{equation}}
\def\f{\end{equation}}
\def\ds{\displaystyle}
\def\_#1{{\bf #1}}
\def\o{\omega}
\def\E{\varepsilon}
\def\M{\mu}
\def\D{\nabla}
\def\.{\cdot}
\def\x{\times}
\def\Re{{\rm Re\mit}}
\def\Im{{\rm Im\mit}}
\def\l#1{\label{eq:#1}}
\def\r#1{(\ref{eq:#1})}
\def\=#1{\overline{\overline #1}}

%

\title{Generalized field-transforming metamaterials}

\author{Sergei A Tretyakov, Igor S Nefedov, and Pekka Alitalo}



\address{Department of Radio Science and Engineering/SMARAD Center of
Excellence,\\ Helsinki University of Technology, P.O. Box 3000,
FI-02015 TKK, Finland} \ead{sergei.tretyakov@tkk.fi}

\begin{abstract}

In this paper we introduce a generalized concept of
field-transforming metamaterials, which perform field transformations
defined as linear relations between the original and transformed
fields. These artificial media change the fields in a prescribed
fashion in the volume occupied by the medium. We show what
electromagnetic properties of transforming medium are required. The
coefficients of these linear functions can be arbitrary scalar
functions of position and frequency, which makes the approach quite general and opens a possibility to
realize various unusual devices.

\end{abstract}

\maketitle

\section{Introduction}

It has been recently realized that metamaterials -- artificial
electromagnetic materials with engineered properties -- can be
designed to control electromagnetic fields in rather general ways.
The concept of ``transformation optics'', which is based on finding artificial
materials that create the desired configuration of electromagnetic fields, has been developed by several
research teams, see e.g.,  \cite{cloak1,Leonhardt,shalaev}

In the known approaches one starts from a certain transformation of
spatial coordinates and possibly also time, which corresponds to
a desired transformation of electromagnetic fields. It has been shown that transformation of spatial
coordinates (accompanied by the corresponding transformation of
electromagnetic fields) can be mimicked by introducing
electromagnetic materials with specific electromagnetic properties
into the domain where the coordinates have been transformed
\cite{cloak1,Leonhardt,leo_cloak}. As an example of such
``coordinate-transforming" device an ``invisibility cloak" has been
suggested \cite{cloak1,leo_cloak}.
In addition to perfect invisibility devices,
similar approaches were applied to perfect
lenses, description of the Aharonov-Bohm effect and artificial black
holes.
It has been known
for a long time that the material relations of an isotropic
magnetodielectric transform into certain bi-anisotropic relations if
the medium is moving with a constant velocity (e.g., \cite{Landau}).
It was suggested that also this effect of time transformation can
possibly be mimicked by a metamaterial \cite{Leonhardt,ab1,biama}.

We have recently introduced an alternative paradigm of transformation
optics, where the required material properties are defined directly
from the desired transformation of electromagnetic fields.
This concept of field-transforming metamaterials was  proposed in
\cite{Roma}, where it was shown what metamaterial properties are
required in order to perform the field transformation of the
form $\_E_0\rightarrow \_E= F(\_r,\omega)\_E_0$,
$\_H_0\rightarrow \_H= G(\_r,\omega)\_H_0$, where $F$ and
$G$ are arbitrary differentiable functions of position and frequency.
In this concept we start directly from a desired
transformation of electromagnetic fields and do not involve any space nor
time transforms.  A
special case of similar transformations was proposed earlier as a numerical
technique  for termination of computational domain by modulating
fields with a function decaying to zero at the termination
\cite{tab}. The physical interpretation of this numerical technique in terms of a slab of a material
with the moving-medium material relations was presented in
\cite{ab1}, and the pulse propagation in a slab of this medium backed by a boundary
was studied in \cite{ab2}.

In this paper we generalize the concept of field-transforming
metamaterials introducing methods to perform general bi-anisotropic
field transformations, where the desired fields depend, in the general linear
fashion, on both electric and magnetic fields of the original field distribution.
Due to generality of the approach, the concept can be applied to
a great variety of field transformations, not limited to cloaking devices.


\section{Field transformations with metamaterials}

Let us assume that in a certain volume $V$ of free space there
exist electromagnetic fields  $\_E_0(\_r)$, $\_H_0(\_r)$ created
by sources located outside volume $V$ (we work in the frequency
domain, and these vectors are complex amplitudes of the fields).
The main idea is to fill volume $V$ with a material in such a way that after
the volume is filled, the original fields $\_E_0$ and $\_H_0$ would be
transformed to other fields, according to the design goals.
Here we consider  metamaterials performing the
general linear field transformation defined as
\e\l{a1} {\bf E(r)} = F({\bf r},\omega){\bf
E_0(r)} + \sqrt{\frac{\mu_0}{\varepsilon_0}}A({\bf r},\omega){\bf
H_0(r)} \f

\e\l{a11} {\bf H(r)} = G({\bf r},\omega){\bf H_0(r)} +
\sqrt{\frac{\varepsilon_0}{\mu_0}}C({\bf r},\omega){\bf E_0(r)}.
\f
Here, scalar functions $F(\_r,\omega)$, $G(\_r,\omega)$
$A(\_r,\omega)$ and $C(\_r,\omega)$ are arbitrary differentiable
functions. In this paper we consider only the case of scalar coefficients in the
above relations, although the method can be extended to the most general
linear relations between original and transformed fields by
replacing scalar coefficients by arbitrary dyadics.

\subsection{Required constitutive parameters}

Substituting \r{a1},\r{a11} into the Maxwell equations and
demanding that the original fields $\_E_0$ and $\_H_0$ satisfy the
free-space Maxwell equations, one finds that the transformed
fields $\_E$ and $\_H$ satisfy Maxwell equations
$\nabla\x\_E=-j\omega \_B$,  $ \nabla\x\_H =j\omega \_D$ in a
medium with the following material relations:

$$
{\bf
B}=\sqrt{\frac{\mu_0}{\varepsilon_0}}\frac{j}{\omega}\frac{1}{FG-AC}(F\nabla
A-A\nabla F)\times {\bf H} + \mu_0\frac{1}{FG-AC}\left(F^2 +
A^2\right){\bf H} $$ \e
 + \frac{j}{\omega}\frac{1}{FG-AC}(G\nabla F-C\nabla A)\times {\bf E}
- \sqrt{\varepsilon_0\mu_0}\frac{1}{FG-AC}\left( AG +
CF\right){\bf E},\l{p1}\f

$${\bf
D}=\sqrt{\frac{\varepsilon_0}{\mu_0}}\frac{j}{\omega}\frac{1}{FG-AC}(C\nabla
G-G\nabla C)\times {\bf E} + \varepsilon_0\frac{1}{FG-AC}\left(G^2
+ C^2\right){\bf E} $$ \e
 + \frac{j}{\omega}\frac{1}{FG-AC}(A\nabla C-F\nabla G)\times {\bf H}
- \sqrt{\varepsilon_0\mu_0}\frac{1}{FG-AC}\left(AG +
 CF \right){\bf H}. \l{p2}\f
Relations \r{p1} and \r{p2} describe a bi-anisotropic medium, whose
constitutive relations can be conveniently written as \cite{biama}

\e {\bf B} = \overline{\overline{\mu}} \cdot {\bf H} +
\sqrt{\epsilon_0\mu_0}(\=\chi+j\=\kappa)^T\cdot {\bf E} , \f

\e {\bf D} = \overline{\overline{\varepsilon}} \cdot {\bf E} +
\sqrt{\epsilon_0\mu_0}(\=\chi-j\=\kappa)\cdot {\bf H}.  \f
Comparing with \r{p1} and \r{p2} we can identify the required material parameters
of the field-transforming medium:

\noindent
The permittivity dyadic
\begin{equation}
\=\varepsilon=\varepsilon_0\frac{1}{FG-AC}\left(G^2 +
C^2\right)\=I +
\sqrt{\frac{\varepsilon_0}{\mu_0}}\frac{j}{\omega}\frac{1}{FG-AC}(C\nabla
G-G\nabla C)\x\=I, \l{eps}\f
the permeability dyadic
\e \=\mu = \mu_0\frac{1}{FG-AC}\left(F^2 +
A^2\right)\=I +
\sqrt{\frac{\mu_0}{\varepsilon_0}}\frac{j}{\omega}\frac{1}{FG-AC}(F\nabla
A-A\nabla F)\x\=I,\f
the nonreciprocity dyadic $\=\chi$
$$ \=\chi = -\frac{1}{FG-AC}(AG+CF)\=I+
$$ \e +\frac{1}{\sqrt{\epsilon_0\mu_0}}\frac{j}{2\omega}\frac{1}{FG-AC}\left[(A\nabla
C-F\nabla G)-(G\nabla F-C\nabla A)\right]\x\=I, \l{chi} \f
and the reciprocal magneto-electric coupling dyadic (also called chirality dyadic) $\=\kappa$
\e
\=\kappa =
\frac{1}{\sqrt{\epsilon_0\mu_0}}\frac{1}{2\omega}\frac{1}{FG-AC}\left[-(A\nabla
C - F\nabla G)-(G\nabla F - C\nabla A) \right]\x\=I. \l{kappa}
\end{equation}
In the above equations  $\=I$ is the unit dyadic.

\section{Classification of generalized field-transforming metamaterials}

In this Section we classify the various cases of different
materials that may need to be obtained depending on the required
transformation~\r{a1},~\r{a11}.

\subsection{Reciprocity and nonreciprocity}

Considering the required permittivity and permeability we see that in the general case
they both contain symmetric and anti-symmetric parts, so the material is
nonreciprocal. The nonreciprocal parts of permittivity and permeability vanish, if coefficients $A$ and $C$
in \r{a1} and \r{a11} equal zero, meaning that the desired
field transformation does not involve magnetoelectric coupling, as in \cite{Roma}. Note that in this case
the medium can be still nonreciprocal due to its magnetoelectric properties, see below.
The other important case when the permittivity and permeability are symmetric (actually
scalar in  the present case) is when the coefficients $F$ and $G$ vanish. These
transformations define new electric field as a function of the original magnetic field and
vice versa.
Furthermore,
we notice that if $A=C=0$ and, in addition,  $G=F$, the wave impedance of the
medium does not change. This is expected, because this is the case when the
electric and magnetic fields are transformed in the same way.

Next, let us consider what types of magnetoelectric coupling in
field-transforming metamaterials are required for various
transformations of fields. Here we will use the general
classification of bi-anisotropic media according to
\cite{biama,class}. This classification is based on splitting the
coupling dyadics $\=\chi$ and $\=\kappa$ into three components,
e.g. for $\=\kappa$ we write

\e \=\kappa = \kappa\=I+ \=N + \=J, \f where
$\overline{\overline{N}}$ and $\overline{\overline{J}}$ denote the
symmetric and antisymmetric parts, respectively.
The scalar coefficient $\kappa$ equals to the trace of $\=\kappa$, and it is
the chirality parameter. In our case it is identically zero, meaning that
field-transforming metamaterials (with scalar transformation
coefficients) are  nonchiral (their microstructure is mirror-symmetric).
Also, the symmetric part $\=N$ is identically zero, and only the
antisymmetric part
\e \=\kappa =
\frac{1}{\sqrt{\epsilon_0\mu_0}}\frac{1}{2\omega}\frac{1}{FG-AC}\left[-(A\nabla
C - F\nabla G)-(G\nabla F - C\nabla A) \right]\x\=I \f
remains.

\begin{table}[h!]
\centering \caption{Classification of reciprocal bi-anisotropic
media.}
\begin{tabular}{|c|c|}
\hline   Coupling parameters & Class \\
\hline $\kappa\neq 0$, $\overline{\overline{N}}=0$, $\overline{\overline{J}}=0$ & Isotropic chiral medium \\
\hline $\kappa\neq 0$, $\overline{\overline{N}}\neq 0$,
$\overline{\overline{J}}=0$ & Anisotropic chiral medium \\ \hline
$\kappa= 0$, $\overline{\overline{N}}\neq 0$,
$\overline{\overline{J}}=0$ & Pseudochiral medium \\ \hline
$\kappa= 0$, $\overline{\overline{N}}=0$,
$\overline{\overline{J}}\neq 0$ & Omega medium \\ \hline
$\kappa\neq 0$, $\overline{\overline{N}}=0$,
$\overline{\overline{J}}\neq 0$ &
Chiral omega medium \\ \hline $\kappa= 0$, $\overline{\overline{N}}\neq 0$, $\overline{\overline{J}}\neq 0$ & Pseudochiral omega medium \\
\hline $\kappa\neq 0$, $\overline{\overline{N}}\neq 0$,
$\overline{\overline{J}}\neq 0$ & General reciprocal
bi-anisotropic medium
\\ \hline
\end{tabular}
\end{table}

Classification of reciprocal bi-anisotropic media is given by Table~1~\cite{biama}.
Field-transforming metamaterials possess reciprocal magneto-electric coupling
as in omega media. The required effects can be realized by
introducing inclusions in form of pairs of the letter $\Omega$, arranging them in
pairs forming ``hats'' \cite{class,radome}.

Similarly, we split the nonreciprocity dyadic  $\=\chi$ as

\e \=\chi = \chi\=I+ \=Q + \=S, \f
where $\overline{\overline{Q}}$
and $\overline{\overline{S}}$ denote the symmetric and
antisymmetric parts, respectively, and $\chi$ is the trace of dyadic $\=\chi$.
From~\r{chi} we can identify
the parts of $\overline{\overline{\chi}}$ as

\e \chi=-\frac{1}{FG-AC}(AG+CF), \f

\e \=Q = 0, \f

\e \=S =
\frac{1}{\sqrt{\epsilon_0\mu_0}}\frac{j}{2\omega}\frac{1}{FG-AC}\left[(A\nabla
C-F\nabla G)-(G\nabla F-C\nabla A)\right]\x\=I. \f
In contrast to the reciprocal magneto-electric coupling, in this case
the trace of the dyadic $\=\chi$ is generally non-zero.
Looking at Table~2~\cite{biama}, we see that the transformation
medium can be either a Tellegen medium, a moving medium, or a
moving Tellegen medium.

\begin{table}[h!]
\centering \caption{Classification of nonreciprocal bi-anisotropic
media.}
\begin{tabular}{|c|c|}
\hline   Coupling parameters & Class \\
\hline $\chi\neq 0$, $\overline{\overline{Q}}=0$, $\overline{\overline{S}}=0$ & Tellegen medium \\
\hline $\chi\neq 0$, $\overline{\overline{Q}}\neq 0$,
$\overline{\overline{S}}=0$ & Anisotropic Tellegen medium \\
\hline $\chi= 0$, $\overline{\overline{Q}}\neq 0$,
$\overline{\overline{S}}=0$ & pseudoTellegen medium \\ \hline
$\chi= 0$, $\overline{\overline{Q}}=0$,
$\overline{\overline{S}}\neq 0$ & Moving medium \\ \hline
$\chi\neq 0$, $\overline{\overline{Q}}=0$,
$\overline{\overline{S}}\neq 0$ &
Moving Tellegen medium \\ \hline $\chi= 0$, $\overline{\overline{Q}}\neq 0$, $\overline{\overline{S}}\neq 0$ & Moving pseudoTellegen medium \\
\hline $\chi\neq 0$, $\overline{\overline{Q}}\neq 0$,
$\overline{\overline{S}}\neq 0$ & Nonreciprocal (nonchiral) medium
\\ \hline
\end{tabular}
\end{table}

If all the transformation coefficients do not depend on the
spatial coordinates, the field-transforming metamaterial is the
Tellegen medium \cite{biama,class}, except the special case when
$AG=-CF$. Tellegen media can be synthesized as composites
containing magnetized ferrite inclusions coupled with small metal
strips or wires \cite{Tellegen} or as mechanically bound particles
having permanent electric and magnetic moments \cite{ChiBi,janus}.

%
%
%
%
%
%
%
%

\subsection{Losses and gain}

A bi-anisotropic medium is lossless if
the permittivity and permeability dyadics are hermittian:
\e \=\varepsilon
=\=\varepsilon^\dagger,\qquad \=\mu=\=\mu^\dagger\f
In addition, the magneto-electric dyadics for lossless media satisfy
\e
\=\chi+j\=\kappa=(\=\chi-j\=\kappa)^*\f
where $*$ denotes complex conjugate \cite{biama}.

Analysing formulas \r{eps}--\r{kappa} for the required material parameters we
see that if the transformation coefficients are complex numbers,
in general the required material can be a lossy or gain-medium. For real-valued
coefficients $F,G,A$, and $C$ the permittivity,
permeability, and reciprocal magnetoelectric coupling coefficient  correspond to lossless media. However,
the antisymmetric part of the nonreciprocity dyadic (describing effects of
``moving'' media) may correspond to lossy or gain media, depending on how the
transformation coefficients depend on the position.

\subsection{Presence or absence of sources}

In the derivation of the material relations in field-transforming
metamaterials \r{p1}, \r{p2} we have assumed that in the medium
there are no source currents, that is, the field equations are
$\nabla\x\_E=-j\omega \_B$, $\nabla\x\_H=j\omega \_D$. In
addition, conditions \e \nabla\cdot \_D=0,\qquad \nabla\cdot
\_B=0\l{sf}\f should be also satisfied in source-free media. For
the general case we have no proof that  conditions \r{sf} always
hold. However, we have checked that they are satisfied in many
important special cases, which include, for example,
position-independent transformation coefficients and exponential
dependence of the coefficients on the position vector.

%
%
%
%
%

\section{Particular cases}

Let us next consider some particular cases of field-transforming metamaterials to
reveal the physical meaning of such transformations of fields.

\subsection{Backward-wave medium}

Consider the case of the field transformation $\_E_0\rightarrow
\sqrt{\mu_0\over\varepsilon_0} \,\_H$, $\_H_0\rightarrow
\sqrt{\varepsilon_0\over\mu_0} \,\_E$, which corresponds to $F=G=0$
and $A=C=1$ in \r{a1},\r{a11}. Transforming electric field
into magnetic field and {\itshape vice versa}, we reverse the
propagation direction of plane waves in the transformation volume,
so we expect that this would correspond to a backward-wave
material filling the volume. And indeed we see that the material
relations \r{p1} and \r{p2} reduce to \e \_B=-\mu_0\_H, \f \e
\_D=-\varepsilon_0\_E,\f which are the material relations of the
Veselago medium \cite{Veselago}.

\subsection{Field rotation}

Consider the case of the field transformation $\_E_0\rightarrow
\sqrt{\mu_0\over\varepsilon_0}\, \_H$, $\_H_0\rightarrow
-\sqrt{\varepsilon_0\over\mu_0}\, \_E$, which corresponds to $F=G=0$
and $A=-1$, $C=1$ in \r{a1},\r{a11}. This kind of
transformation relates to changing the polarization of the field:
fields of a plane wave are rotated
by $-90$ degrees around the propagation axis. The material relations \r{p1} and \r{p2} reduce to
\e \_B=\mu_0\_H, \f \e \_D=\varepsilon_0\_E,\f
which are the
material relations of free space. The reason why we do not see the
change of polarization here is simply due to the fact that we
wanted to consider $F$, $G$, $A$ and $C$ as \textit{scalars}. The
media (original or transformed) do not ``feel'' the polarization of
the fields, so the rotation does not have any influence on the
material relations. For instance, unpolarized light is obviously invariant under this
transformation. We want to remind the reader that by considering
the transformation coefficients $F,G,A,C$ as dyadics, one can obtain
more general transformations, which tailor also the field polarization in the
general way.

\subsection{Isotropic Tellegen medium}

Consider next a more general case when all the transformation
coefficients do not depend on the position vector inside the
transformation domain. In this case, gradients of these functions
vanish, and we see that the material relations are that of an
isotropic Tellegen material: \e \_B=\mu_0\frac{1}{FG-AC}\left(F^2
+ A^2\right)\_H-\sqrt{\varepsilon_0\mu_0}\,\frac{AG+CF}{FG-AC}\,\_E, \f \e
\_D=\varepsilon_0\frac{1}{FG-AC}\left(G^2 +
C^2\right)\_E-\sqrt{\varepsilon_0\mu_0}\,\frac{AG+CF}{FG-AC}\,\_H. \f The medium is
reciprocal only if $\chi = -\frac{AG+CF}{FG-AC}=0$. We can note that if the transformation coefficients vary
within the transformation volume, this generally requires
simulation of a moving medium with the transformation
metamaterial.

\subsection{Tellegen nihility}

Let us consider again the case where all the coefficients are
position-independent, that is, \e \nabla F = \nabla G = \nabla A = \nabla C =0 \f
and also demand that the permittivity and permeability of the transforming medium become zero.
This takes place when either
\e A=jF,\qquad C=jG\l{plus}\f
or
\e A=-jF,\qquad C=-jG\l{minus}\f
The material relations of the medium become

\e\l{d0}\begin{array}{l} {\bf
B}=-j\sqrt{\varepsilon_0\mu_0}\, {\bf E} \\ {\bf
D}=-j\sqrt{\varepsilon_0\mu_0}\, {\bf H}\end{array}\f
for the case \r{plus} or
\e\l{d1}\begin{array}{l} {\bf
B}=j\sqrt{\varepsilon_0\mu_0}\, {\bf E} \\ {\bf
D}=j\sqrt{\varepsilon_0\mu_0}\, {\bf H}\end{array}\f
for the case \r{minus}.

These constitutive relations look similar to that of chiral nihility \cite{CN},
but in this case the only non-zero material parameter is the Tellegen parameter.
The Maxwell equations corresponding to relations \r{d1} take the form
\e\l{d11}\begin{array}{ll} \nabla\times{\bf
H}=-k_0{\bf H}, & \nabla\times{\bf
E}=k_0{\bf E},\end{array}\f
where $k_0=\omega\sqrt{\varepsilon_0\mu_0}$ is the free-space wavenumber.
Thus, solutions of Eq.~\r{d11} are eigenfunctions of operator rot=$\nabla\times\=I$.
If $\partial/\partial x=\partial/\partial y=0$ then for the waves propagating in $z$-direction
\e\l{d0}\begin{array}{ll}
E_x=-e_0\exp{(-jk_0z)} & H_x=e_0\eta^{-1}\exp{(-jk_0z)}\\
E_y=je_0\exp{(-jk_0z)} & H_y=je_0\eta^{-1}\exp{(-jk_0z)}\\
E_z=0  &  H_x=0
\end{array}\f
where $\eta$ is the wave impedance. Substituting expressions \r{plus} or \r{minus} for $A$ and $C$ into the
original definitions \r{a1},\r{a11}, we obtain the following formula for the wave impedance:
\e\l{d4}
\eta=\frac{F}{G}\eta_0
\f
where $\eta_0$ is the wave impedance of vacuum.

As in the case of chiral nihility \cite{CN} and the Beltrami fields in chiral media \cite{ChiBi,Beltr},
the fields \r{d0} correspond to circularly-polarized waves. However,
in contrast to those two cases, where the phase shift between
electric and magnetic fields equals exactly to $\pi/2$, in the considered case it
depends on relation between $F$ and $G$ according to \r{d4}. If $F=G$, the electric and magnetic fields oscillate in phase.

\subsection{Position-independent transformation coefficients}

An interesting general conclusion can be drawn with regard to field transformations
with all the coefficients being position-independent. In this case
the wavenumber in the field-transforming metamaterial is the same as in the original
medium (free space in our case).  This can be shown by
using the equation for the wavenumber $k$ in a Tellegen
medium~\cite{ChiBi}

\e k=k_0\sqrt{\varepsilon_r\mu_r-\chi^2},
\l{Tellegen1} \f
where $\varepsilon_r$ and $\mu_r$ are the relative permittivity and permeability.
Since \e \varepsilon_r\mu_r-\chi^2 =
\frac{(G^2+C^2)(F^2+A^2)}{(FG-AC)^2}-\frac{(AG+CF)^2}{(FG-AC)^2}=1,
\f \r{Tellegen1} simplifies to \e k=k_0. \f
This means that for transformations of this class only the wave impedance of the
transforming medium needs to be changed, but not the wavenumber. One can notice that
this property indeed holds in all of the above special cases where the
coefficients are position-independent.

\subsection{Moving omega medium}

Next we consider another class of field-transforming media assuming
$A({\bf r},\omega)=C({\bf r},\omega)=0$. Transformations of this
type correspond to moving omega materials with the constitutive
relations of the form:
\begin{equation}\begin{array}{ll}
\_D=\varepsilon_0\frac{G}{F}\_E-\frac{j}{\omega G}\nabla G\times
\_H, & \_B=\mu_0\frac{F}{G}\_H+\frac{j}{\omega F}\nabla F\times
\_E.\end{array}
  \l{b1}\end{equation}
Both the nonreciprocity dyadic $\=\chi$ and the reciprocal
magnetoelectric coupling dyadic $\=\kappa$ are antisymmetric:
\begin{equation}
\=\chi=-\frac{j}{2k_0}\left(\frac{\nabla F}{F}+\frac{\nabla
G}{G}\right)\times\=I ,\qquad
\=\kappa=-\frac{1}{2k_0}\left(\frac{\nabla F}{F}-\frac{\nabla
G}{G}\right)\times\=I, \label{j1}
 \end{equation}
where the first quantity describes moving-media effects and the
second one corresponds to omega-medium properties. If the two
transformation coefficients are equal ($F=G$), the omega coupling vanishes.

The Maxwell equation for the transformed fields read:
\begin{equation}\begin{array}{l}
\nabla\times\_E(\_r)=-j\omega\mu_0\frac{F(\_r)}{G(\_r)}\_H(\_r)+\frac{1}{F(\_r)}\nabla
F(\_r)\times\_E(\_r), \\
\nabla\times\_H(\_r)=j\omega\epsilon_0\frac{G(\_r)}{F(\_r)}\_E(\_r)+\frac{1}{G(\_r)}\nabla
G(\_r)\times \_H(\_r).
 \end{array}
  \label{a2}\end{equation}

\subsection{Artificial moving medium}
\label{amm}

Let us consider a more special case when the transformation preserves the
wave impedance of the medium  [$F(\_r)=G(\_r)$] \cite{Roma}. Material relations
\r{b1} take the form of material relations of slowly moving media (velocity $v\ll c$,
where $c$ is the speed of light in vacuum), see, e.g., \cite{Kong}:
\e
\begin{array}{ll}\_D={1\over\sqrt{1-v^2/c^2}}\left(\epsilon_0\_E+{1\over c^2}\_v\x\_H\right)\approx \epsilon_0\_E+{1\over c^2}\_v\x\_H\\
\_B={1\over\sqrt{1-v^2/c^2}}\left(\mu_0\_H-{1\over c^2}\_v\x\_E\right)\approx\mu_0\_H-{1\over c^2}\_v\x\_E
\end{array} \l{moving}\f
In the last relation we have neglected the second-order terms $v^2/c^2$. The ``velocity'' of the
transforming medium can be identified by comparing with \r{b1} as
\e \_v=-j{c^2\over \omega}{\nabla F\over F}\f

To get more insight, let us analyze the
one-dimensional case considering transformation functions depending only on one
coordinate  $z$ and assuming that $F(z)=G(z)$ is an exponential function of $z$. Two cases should be distinguished:

\begin{itemize}
\item $\nabla F/ F$ is {\itshape imaginary}, e.g., $F(z)=e^{-j\alpha z}$
with a real parameter $\alpha$. In this case the effective ``velocity" is {\bf real}.
This metamaterial simulates moving media. Note that the required material relations
 have the form of the relations for slowly moving media even if the equivalent velocity is not small.

\item
 $F(z)$ is a purely {\itshape real} function, e.g., $F(z)=e^{-\alpha z}$,
 where $\alpha$ is real. In this case  the effective ``velocity" of the medium is {\bf imaginary}.
Note that the material relations of a medium traveling faster than light formally
have an imaginary vector coefficient  in the second term of \r{moving}. However, one cannot
say  that this metamaterial simulates media moving faster than light, because in that case also the
effective permittivity and permeability would be imaginary quantities.

\end{itemize}

Let us examine the source-free conditions $\nabla\cdot {\bf D}=0$ and $\nabla\cdot {\bf B}=0$ for this special
case of $F=G=e^{-\alpha z}$. Actually,
$\frac{\nabla F(\_r)}{F(\_r)}=\frac{\nabla G(\_r)}{G(\_r)}=-\alpha\,\_z_0$. Then it follows from \r{b1}
 \e \nabla\cdot\_D=\nabla\cdot[F(\_r)\epsilon_0\_E_0(\_r)]+ \frac{j}{\omega}\nabla\cdot[\alpha\_z_0\times F(\_r)\_H_0]\f
Simple calculation using the free-space Maxwell equations relating $\_E_0$ and $\_H_0$
shows that $\nabla\cdot {\bf D}=0$.
Condition $\nabla\cdot{\bf B}=0$ can be checked similarly.

Let us check the passivity condition for this medium.
The period-average of the time derivative of the field energy is
expressed as \cite{biama}:
\begin{equation}
\left\langle\frac{dW}{dt}\right\rangle_t=\frac{j\omega}{4}{\rm
f^*\cdot(M-M^\dagger)\cdot f}
\end{equation}
where
\begin{equation}\begin{array}{ll}
{\rm f}=\left(\begin{array}{l} \_E \\ \_H\end{array}\right), &
 {\rm M}=\left(\begin{array}{ll} \epsilon_0 {\overline{\overline I}}  & \sqrt{\epsilon_0\mu_0}\,{\overline{\overline \chi}}^\dagger\\
\sqrt{\epsilon_0\mu_0}\,{\overline{\overline \chi}} & \mu_0
{\overline{\overline I}}
\end{array}\right).
\end{array}\end{equation}
For real transformation coefficients we take as an example  $F(z)=e^{-\alpha z}$,
where $\alpha>0$.
For a plane wave propagating along $z$, calculation leads to
\begin{equation}
\left\langle\frac{dW}{dt}\right\rangle_t=\frac{\alpha}{\eta}e^{-2\alpha
z}>0\end{equation}
This implies that the medium possesses losses.
However, for a plane wave traveling in the opposite direction (along $-z$),
the sign in the above expression is reversed, corresponding to a gain medium.
This is expected because in the last case the field amplitude grows along the
propagation direction.
In the case of imaginary $\alpha$
this value is equal to zero and the medium is lossless.

\section{Transmission through a slab of a field-transforming medium}

Let us consider an infinite slab of a field-transforming metamaterial in vacuum, which is excited
by a normally incident plane wave. Here we continue studying the case of
moving-medium transformations, that is, $F=G$ and $A=C=0$. Let the thickness of the slab be $d$, and
the field-transforming properties be defined by a function of one
variable $z$ --- the coordinate in the direction normal to the interfaces: $F(z)=G(z)$.
The general solution for the eigenwaves in the slab is, obviously,
\e E(z)=F(z)(a e^{-jk_0z}+b e^{jk_0z})\f
where $E$ is the electric field amplitude, and $a$ and $b$ are constants,
because this is how the free-space solution is transformed in this metamaterial slab.
If $F=G$, the wave impedance in the transforming medium does not change.

Writing the boundary conditions on two interfaces, reflection and transmission coefficients
can be found in the usual way. As a result, we have found that the reflection coefficient
does not depend on function $F(z)$ at all, and it is given by the standard
formula for a reflection coefficient from an isotropic slab at normal incidence. In
particular, if the permittivity and permeability of the slab equal to that of
free space, the reflection coefficient $R=0$. The transmission coefficient in this
matched case reads
\e T={F(d)\over F(0)}\, e^{-jk_0d}\f
where $F(d)$ and  $F(0)$ are the values of the transformation function at the
two sides of the slab.

The non-reciprocal nature of field-transforming metamaterials is clearly visible here.
If function $F(z)$ is real and it decays with $z$, then fields of waves traveling in the
positive direction of $z$ decay, while fields of waves traveling in the opposite
directions grow. Obviously, such performance cannot be realized by passive media (see the
calculation of the absorbed power in Section~\ref{amm}). If $F(d)=F(0)=1$, the slab is
``invisible'', although inside the slab  the field distribution can be pretty arbitrary,
as dictated by the given function $F(z)$.

\section{Conclusions}

Two different approaches to design of metamaterials which transform fields in a desired
way are known.
The first one has been proposed and developed in \cite{cloak1,Leonhardt}, and it is
based on coordinate transformations. The second one has been introduced in \cite{Roma} and
developed here, and it transforms
the fields directly, according to a desired prescription.
This technique
can be potentially applied to a very
 wide variety of engineering problems, where a certain field
 transformation is required.
One of the most important property of the field transformation
approach is that a medium with prescribed parameters turns out to
be nonreciprocal with except of some special cases. Most often, a
metamaterial simulating moving media is required. Potential
physical realizations of metamaterials with the properties
required for general linear field transformations (such as
artificial moving media) are discussed in
\cite{ab1,biama,Tellegen,janus,Zagr}.


\section*{Acknowledgements}

This work has been partially funded by the Academy of Finland and
TEKES through the Center-of-Excellence program and partially by
the European Space Agency (ESA--ESTEC) contract no. 21261/07/NL/CB
(Ariadna program). The authors wish to thank prof. Constantin
Simovski for helpful discussions. P.~Alitalo acknowledges
financial support by the Finnish Graduate School in Electronics,
Telecommunications, and Automation (GETA), the Emil Aaltonen
Foundation, and the Nokia Foundation.

\end{document}